
%
%
%
%
\input harvmac.tex
\def\t{\theta}
\def\e{\epsilon}
\def\ep{\epsilon}
\def\<{\langle}
\def\>{\rangle}
\noblackbox
\Title{\vbox{\baselineskip12pt
\hbox{USC-93-022}}}
{\vbox{\centerline{Massless integrable quantum field theories}
\vskip2pt\centerline{ and massless scattering in 1+1 dimensions
}}}

\centerline{P. Fendley$^\dagger$ and H. Saleur$^{\spadesuit *}$}
\vskip2pt
\centerline{$^\dagger$Department of Physics, University of Southern California}
\centerline{Los Angeles CA 90089}
\vskip2pt
\centerline{$^\spadesuit$Department of Physics and Department of Mathematics}
\centerline{University of Southern California}
\centerline{Los Angeles CA 90089}
\vskip.3in

These lecture notes provide an elementary introduction to the study
of massless integrable quantum field theory in 1+1 dimensions using
``massless scattering''.  Some previously unpublished results are also
presented, including a non-perturbative study of Virasoro conserved
quantities.
\bigskip
\bigskip\centerline{\it Lectures presented at Strings 93, May 1993}
\centerline{\it and at the Trieste Summer School
on High Energy Physics and Cosmology, July 1993.}
\bigskip\bigskip
\noindent $^*$ Packard Fellow
\Date{9/93}

\newsec{Introduction}

A massless quantum field theory has no gap in the excitation spectrum.
This can be seen, for example, in the Laplace representation of
Green's functions. This of course does not imply scale invariance; in
general properties will interpolate between those of the two different
conformal field theories describing the UV and IR fixed points. A
standard (although a little marginal) example of such theory is the
$O(3)$ non-linear sigma model with topological angle $\Theta=\pi$,
which has central charges $c_{UV}=2$ and $c_{IR}=1$.

\nref\AFL{N. Andrei, K. Furuya, and J. Lowenstein, Rev. Mod. Phys. 55
(1983) 331; A.M. Tsvelick and P.B. Wiegmann, Adv. Phys. 32 (1983) 453.}
\nref\aff{I. Affleck, in Les Houches 1988 {\it Fields, Strings, Critical
Phenomena}, ed. by E. Brezin and J. Zinn-Justin, North Holland.} In 1+1
dimensions, many massless theories are integrable.
Such theories include well-known statistical-mechanical models like
the continuum limit of the XXZ spin chain and the Kondo problem. Many more are
provided by
appropriate perturbations of conformal field theories.  Their study is
interesting for several reasons. A few properties are accessible
experimentally; see for example \refs{\AFL,\aff}.  Features of
academic interest include Green's functions with different anomalous
dimensions in the UV and IR, the consequences for the topology of the
space of relativistic quantum field theories, a better understanding
of the second law of thermodynamics associated with renormalization
group trajectories, and a way of understanding perturbations of IR
fixed points by irrelevant operators.

\nref\FT{L.D. Faddeev and L.A. Takhtajan, Phys. Lett.  85A (1981) 375.}
\nref\S{F. Smirnov, Th. Math. Phys.  60
(1984) 363.}
\nref\KII{V.E. Korepin, Comm. Math. Phys. 86 (1982) 391.}
In these lectures we will discuss the scattering theories associated
with integrable massless quantum field theories.  In a massless theory
the excitations should consist of right-moving and left-moving
particles with $p=\pm E$, where we set the speed of light to be 1.
$S$-matrices describing the ``scattering'' of such particles were
calculated long ago in \FT\ for the XXX model.
Such objects do not make much sense in traditional $S$-matrix theory
where one requires the existence of in and out states; it is difficult
for instance to imagine a physical process that would lead to
scattering between two particles moving in the same direction at the
speed of light. Massless $S$-matrices in $1+1$ dimensions seem to make
sense only in the context of integrable quantum field theories.  In
this case the scattering is completely elastic: momenta are conserved
individually.  We build states by acting with creation and
annihilation operators on the ground state. These operators have
non-trivial commutation properties (the Zamolodchikov-Faddeev algebra
\refs{\S,\KII}) encoded in the $S$-matrix (suppressing internal
indices describing the particles):
\eqn\comm{{\cal R}^+(\theta_1){\cal R}^+(\theta_2)
=S(\theta_1-\theta_2){\cal R}^+ (\theta_2){\cal R}^+(\theta_1),}
where by convention ${\cal R}^+(\theta_i){\cal R}^+(\theta_j)$ creates
a plane wave with $x_i<x_j$.  This formal definition of $S$, directly
inspired by the Bethe ansatz equations, makes sense in both massive
and massless cases. Another way of describing this definition is as a
matching condition on two-particle wavefunctions.

Some care must be taken when defining massless integrable theories.
For instance, the usual proof that an infinite number of conserved
quantities implies factorized scattering relies on the possibility of
separating wave packets in general \ref\P{S. Parke, Nucl. Phys. B177
(1980) 166.}, which is not possible in the massless case. Analyticity
properties are also not completely clear.  Since the particles are
massless, they are either right- or left-moving. Because $S$-matrices
for pure left-left or right-right scattering can be obtained by a
limiting process from physical $S$-matrices acting on massive
particles, one requires from $S_{LL}$ and $S_{RR}$ exactly the same
properties as for physical $S$-matrices. As we will discuss, the case
of $S_{LR}$ is subtler. In general these properties can be derived
using the definition based on Bethe ansatz wavefunctions, without
reference to any in and out states.

After having given the caveats, we would like to explain why
finding massless $S$-matrices is a worthy endeavor.
Even if the $S$-matrix found has no physical meaning in its own right,
many quantities calculated from it do. In these notes we will show how
to calculate the free energy at non-zero temperature using the
thermodynamic Bethe ansatz (TBA). Modular invariance relates this to
the Casimir energy on a circle, giving a ``c-function'', which
for unitary models shows the evolution of the number of degrees of
freedom in the flow from ultraviolet to infrared.  This calculation
can also be modified to give some excited-state energies as well,
which give conformal dimensions in the critical limit. We will also
show how to obtain the ground-state energy at zero temperature in a
background field, a result related to the chiral $U(1)$ anomaly in the
critical limit.  Finally, we present here some new results on
obtaining higher-spin Virasoro conserved charges from the massless
scattering.

These results in fact suggest that some aspects of a conformal field
theory can be described by a theory of massless particles with no
left-right scattering \ref\ZZI{A.B. Zamolodchikov, Al.B.
Zamolodchikov, Nucl. Phys.  B379 (1992) 602.}. This can be seen as
follows.  In these massless but not scale-invariant theories, we have
a mass scale $M$; $M=0$ gives the ultraviolet fixed point while
$M\rightarrow\infty$ gives the infared one. The momenta and energy of
the particles are parametrized by
\eqn\spec{\eqalign{
E=p&={M\over 2}e^\theta\ \ \ \ \ \ \ \ \ \ \ \hbox{for right movers}\cr
E=-p&={M\over 2}e^{-\theta}\ \ \ \ \ \ \ \ \ \hbox{for left movers}}}
A Lorentz-invariant $S$-matrix element $S_{LL}$ describing scattering
of two left movers depends only on the ratio of the two momenta, so it
depends only on $\t_1-\t_2$ and not on $M$. The left-right scattering
also depends only on the rapidity difference but does depend on $M$,
because the only Lorentz invariant is $s=(p_1+p_2)^2$.  We can always
rescale $M\rightarrow\infty$ by shifting the rapidities. The $LL$ and
$RR$ $S$-matrices are independent of this shift (although $S_{LR}$ is
not), so they are characterized solely by properties of the infrared
fixed point. In this sense one can think of the $LL$ and $RR$
$S$-matrices as being the $S$-matrices for the conformal field theory.
This should not seem bizarre --- many properties of four-dimensional
field theories (even ones with massless particles like QED) are
described by particle theories!

\nref\ZZ{A.B. Zamolodchikov and Al.B. Zamolodchikov, Ann. Phys. 120
(1979) 253.}
\nref\ZAdv{A.B. Zamolodchikov, Adv. Stud. Pure Math. 19 (1989) 1. }
\nref\Muss{G. Mussardo, Phys. Rep. 218 (1992) 215.}
Another reason for studying massless $S$-matrices is that finding them
is often an easier task than doing the full Bethe ansatz, a result of
the constraints of an integrable theory. In the massive case, this has
become a highly-developed art (see \refs{\ZZ-\Muss} for reviews), and
many of these lessons can be applied to the massless case. The basic
method is to guess the particle content based on the knowledge of the
symmetries of the problem (and on the Lagrangian, if one is known),
and then find the simplest $S$-matrix consistent with these symmetries
as well as the criteria of factorizability, unitarity and crossing
symmetry. In many cases, such an $S$-matrix is the correct one, as can
be checked by a variety of methods. The symmetries, in particular
affine quantum group symmetries, must however be analyzed carefully.

\nref\ZI{Al.B. Zamolodchikov, Nucl. Phys. B358 (1991) 524.}
\nref\FSZ{P. Fendley, H. Saleur and Al.B. Zamolodchikov, ``Massless Flows I''
and ``Massless Flows II'',
hepth \#9304050 and 9304051, to appear in Int. J. Mod. Phys. A}
\nref\pkon{P. Fendley, Phys. Rev. Lett. 71 (1993) 2485.}
\nref\sausage{V. Fateev, E. Onofri and Al.B. Zamolodchikov, ``The
sausage model (integrable deformations of O(3) sigma model)'', to appear
in Nucl. Phys. B.}
\nref\pk{P. Fendley and K. Intriligator,
``Exact $N$=2 Lan\-dau-Ginz\-burg Flows'', hepth \#9307166, to appear
in Nucl. Phys. B.}
\nref\rjpa{N. Reshetikhin, J. Phys. A24 (1991) 3299.}
\nref\chbo{J. Carmelo, P. Horsch, P.A. Bares and A.A. Ovchinnikov,
Phys. Rev. B44 (1991) 9967.}
\nref\esskor{F. Essler and V. Korepin, ``Scattering Matrix
and Excitation Spectrum of the Hubbard Model'', ITP-SB-93-40.}
In addition to the XXX model, massless $S$-matrices have been found for a
number of models. Continuum theories  include the flow from
the tricritical Ising model to the Ising model \ZI, the
$O(3)$ sigma model at $\Theta=\pi$ and the $SU(2)_1$ principal chiral
model \ZZI, the flows between the minimal models \FSZ, the Kondo
problem \pkon, the ``sausage'' sigma model \sausage, and the
Landau-Ginzburg flows to the $N$=2 minimal models \pk.  Lattice
models include integrable higher-spin XXX chains \rjpa\ and the
Hubbard model \refs{\chbo,\esskor}.

The purpose of these lectures is to provide a pedagogical
introduction, so we will skip many technical details.  We have tried
to make the sections reasonably independent of one another so that
they can be read separately. In Section 2 we give a simple example of
a massless field theory, the sine-Gordon model with imaginary
potential.  Section 3 contains a discussion of the usual Thirring
model (for properties we discuss, Thirring and sine-Gordon can
be used interchangeably) and its massless limit. Its main purpose is to
introduce the physically odd idea of left-left or right-right
scattering between massless particles from the Bethe ansatz view
point. Section 4 gives a simple introduction to the thermodynamic
Bethe ansatz carried out with the example of massless scattering.
Section 5 explores a little more how one can describe certain
aspects of conformal field theories by massless particles with no
left-right scattering.  Previously unpublished results regarding the
non-perturbative analysis of conserved quantities in conformal field
theories are presented.  Section 6 contains some comments on
integrable theories with nontrivial left-right scattering. Section 7
is based on \FSZ. There we carry out explicitly the study of the
sine-Gordon model with imaginary coupling and background field, the
latter being introduced to get a simpler calculation. We show that
even if massless scattering appears a little odd physically, it at
least provides the proper analytic continuation of physical quantities
in appropriate directions of the parameter space. Section 8 contains
conclusions and questions of interest.

\newsec{A simple example of  massless field theory}

A simple but generic example of integrable massless field theory is
provided by the sine-Gordon model with imaginary ``coupling'' (i.e.\
prefactor of the cosine term). Recall the hamiltonian
\eqn\sgsg{H=\int ds\left[{1\over 2}\Pi^2+{1\over 2}(\partial\phi)^2+
\lambda\cos\beta_{SG}\phi\right]}
where we set $\beta_{UV}^2=8\pi{t\over t+1}$ at the UV fixed point.
This model is well-studied for $\lambda$ real and known to be a
massive field theory with a trivial fixed point in the IR. Taking
$\lambda$ imaginary does not look too physical at first. There are
however several reasons for doing so. The massless flows between
minimal models are obtained as reductions of this model, and are
unitary even though \sgsg\ is not. In addition, \sgsg\ describes the
flow of the $O(n)$ model to its low-temperature phase
\ref\N{B. Nienhuis, Phys. Rev. Lett. 49 (1982) 1062.}; this covers
interesting physical situations in condensed-matter physics such as
self-avoiding polymers.

To see the difference in behavior between $\lambda$ real and imaginary
consider the large-$t$ limit, where the cosine term is almost marginal
and reliable perturbation theory at order $1/t$ can be carried out
following the analysis of the XY model \ref\JKKN{J.V. Jose, L.P.
Kadanoff, S. Kirkpatrick, D.R. Nelson, Phys. Rev. B16 (1977) 217.}.
The RG equations are
\eqn\tata{{d\lambda\over d\beta}=\left(2-{\beta^2\over 4\pi}\right)\lambda,}
and
\eqn\tataI{{1\over\beta^2}{d\beta^2\over db}=-\pi^2\lambda^2.}
{}From these one deduces, at first nontrivial order
\eqn\tataII{{d\beta^2\over db}=-{\beta^4\over 4\pi}\left[
\left({8\pi\over\beta^2}-1\right)^2-{1\over t^2}\right].}
For $\lambda$ real, the initial derivative of $\beta$ is negative, and
the coupling flows to zero at large distance. For $\lambda$ imaginary,
the initial derivative is positive and the coupling increases
monotonically to the IR fixed point with
\eqn\tataIII{\beta_{IR}^2=8\pi{t-1\over t}}
Correspondingly both the UV and IR fixed points are Gaussian models with
different radii of compactification, and we have a flow ``within'' $c=1$.

The most interesting aspect concerns the evolution of the running
central charge. The latter can be defined in several ways away from
the fixed points, for instance using the two-point function of the
stress energy tensor \ref\ZII{A.B. Zamolodchikov, JETP Lett. 43 (1986)
730.} or finite-size effects\nref\IS{C. Itzykson and H. Saleur, J. Stat.
Phys.  48 (1987) 449.}\nref\LC{A. Ludwig and J. Cardy, Nucl. Phys. B285
(1987) 687.} \refs{\IS,\LC}. Qualitatively, these functions should
behave in a similar way and are usually expected to describe the
evolution of the number of degrees of freedom. By analogy with
the second law of
thermodynamics we expect that  such a function should decrease when
following a RG flow. It has been proven that the first type of
$c$-function always decreases in unitary theories (the $c$-theorem
\ZII), and all known unitary examples of the second type also
decrease.  In our non-unitary problem, it is easy to compute them at
first non trivial order in $1/t$ where they coincide.  One finds then
\FSZ
\eqn\crun{c=1+{12\over t^3}{e^{4(b-b_0)/t}
(1-e^{4(b-b_0)/t})\over (1+e^{4(b-b_0)/t})^3},}
that has a roaming behavior. It indeed has $c=1$ in the UV and IR but
goes up and down in between, reaching a pair of extrema with values
\eqn\cext{c_{\pm}=1\pm{2\over\sqrt{3}}{1\over t^3}.}
Of course the fact that $c$ can increase and does not obey a
nonunitary version of the $c$-theorem is not surprising and is an
obvious consequence of the imaginary coupling.  One might hope to
slightly modify the perturbation to obtain an exact flow which would
stop at the value $c_+$. This does not seem impossible in view of
\ref\Zo{Al.B. Zamolodchikov, ``Resonance factorized scattering
and roaming trajectories'', Ecole Normale preprint ENS-LPS-355}, which
describes flows that exactly interpolate between minimal models and
flows that go ``very near'' them.

The fact that $c$ is not monotonic sheds some doubt on the
reliability of the running $c$-function as a measure of the number of
degrees of freedom. Let us emphasize that \crun\ is indeed related to
the absolute ground state of the theory --- it {\it is} the ``$c_{eff}$''
supposed to qualitatively replace $c$ in the non-unitary case.

\newsec{Massless scattering at the conformal point}
\nref\BT{H. Bergknoff and H.B. Thacker, Phys. Rev. D19 (1979) 366.}
\nref\Kor{V.E. Korepin, Th. Math. Phys. 41 (1979) 953.}

We consider the example of the massive Thirring model on a circle of
length $L$ with hamiltonian
\eqn\thir{H=\int dx \left[-i\left(\psi_1^+\partial_x\psi_1-\psi_2^+
\partial_x\psi_2\right)+m_0\left(\psi_1^+\psi_2+\psi_2^+
\psi_1\right)+2g_0\psi_1^+\psi_2^+\psi_2\psi_1\right].}
which is solvable by the Bethe ansatz \refs{\BT,\Kor}. The analogous
calculation is done for the sine-Gordon model in \ref\FST{L. Faddeev,
E. Sklyanin and L. Takhtajan, Th. Math. Phys. 40 (1979) 688.}. As a
result eigenenergies take the form
\eqn\en{{\cal E}=\sum_i m_0\cosh\xi_i,}
and  momenta
\eqn\ppp{P=\sum_i m_0\sinh\xi_i,}
where the $\xi_i$ are bare rapidities of pseudoparticles satisfying
the Bethe ansatz equations
\eqn\ba{\exp(im_0L\sinh\xi_i)\prod_j{\sinh(\xi_i-\xi_j+2i\mu)\over\sinh(\xi_i
-\xi_j-2i\mu)}=1,}
and $\cot\mu=-{1\over 2}g_0$. One recognizes in \ba\ the conditions
for the wave function to be periodic, the phase shift being a
combination of a free term and factorized scattering between pairs of
pseudoparticles. Thus the elements of the product in \ba\ are the ``bare''
$S$-matrix elements for the pseudoparticles, which we denote as
$S_0(\xi_i-\xi_j)\equiv\exp[i\phi_0(\xi_i-\xi_j)]\equiv{1+i\Lambda\over
1-i\Lambda}$.  Equation \ba\ follows from the form of Bethe wave
functions
\eqn\wf{\psi(x_1,\ldots,x_N|\xi_1\ldots,\xi_N)=
\exp(im_0\sum_i x_i\sinh\xi_i)\prod_{i<j}[1+
i\Lambda(\xi_i-\xi_j)\hbox{sign}(\xi_i-\xi_j)].}
This wave function is almost free, with only a phase shift when the
coordinates are exchanged. Of course since we are dealing with
fermions, the real wave function has still to be antisymmetrized so
\eqn\anti{\Psi(x_1,\ldots,x_N)=\sum_P\hbox{sign}(P)
\psi(x_1,\ldots,x_N|\xi_{P1},\ldots,\xi_{PN}),}
(this makes sense because $\Lambda$ is an odd function, or
equivalently $S_0(\xi)S_0(-\xi)=1$) and all rapidities must therefore
be different.

The study of the quantum field theory requires building the ground
state by filling the appropriate Dirac sea, and then finding the
excitations and their scattering.  We must mention a difficulty
concerning the choice of UV cutoff. In the attractive regime $g_0>0$
essentially all cutoffs produce the same results, but the situation is
different in the repulsive regime $g_0<0$. (This is unfortunately
the regime where one truncates the sine-Gordon model to describe
massive perturbations of Virasoro minimal models by the $\phi_{13}$
operator
\ref\EY{T.  Eguchi and S.K. Yang, Phys. Lett.  B224 (1989) 373; T.
Hollowood and P. Mansfield, Phys. Lett. 226B (1989) 73; M.T. Grisaru,
A. Lerda, S. Penati and D. Zanon, Phys. Lett. B234 (1990) 88; N.
Reshetikhin and F. Smirnov, Comm. Math. Phys. 131 (1990) 157.}.) The
rapidity cutoff of
\ref\KI{V.  Korepin, Comm. Math.  Phys. 76 (1980) 165.} leads to many
more particles in the spectrum than the cutoff of
\ref\W{G. Japradize, A. Nersesyan and P. Wiegmann, Nucl. Phys. B230 (1984)
511; P. Wiegmann, Phys. Lett. B152 (1985) 209.}, which is essentially
a lattice cutoff \ref\RS{N.  Reshetikhin and H. Saleur, ``Lattice
regularization of massive and massless integrable field theories'',
preprint USC-93-020, hepth \#9309135.}. The latter regularization
reproduces the conjectures of
\ZZ\ and seems in agreement with the expected physics of perturbed
minimal models \ref\FS{P. Fendley and H. Saleur, Nucl. Phys.  B388 (1992)
609.}. We avoid entering into technical details and give a ``morally''
correct discussion in the following.

Another difficulty arises in the eigenfunctions of \thir\ as a result of
terms like ($\hbox{sign}\ \delta$).  Depending on the regularization,
different relations between the bare coupling $g_0$ and the parameter
$\mu$ are found.  In \Kor\ the relation $\mu={\pi+g_0\over 2}$ is used
instead. We shall carry out the discussion in terms of the variable
$\mu$.

The ground state is easily built by filling the sea with
antipseudoparticles, i.e.\ filling the line $Im(\xi)=\pi$. There are
various kind of excitations.  We shall discuss only the case of
solitons $s$ and antisolitons $a$ (these are the only particles with
non-vanishing $U(1)$ charge).  For instance a pair of solitons is
simply obtained by making two holes in the ground state distribution.
Antisolitons, or pairs of solitons and antisolitons are obtained
similarly, with the addition of some strings \Kor\ of pseudoparticles
around $Im(\xi)=0$. Of course the introduction of such holes induces
a shift in the distribution of the $\xi$'s, the so-called {\it
backflow}. As a result the mass and rapidity of the
solitons are renormalized, giving
\eqn\mp{E=m\cosh\theta,\ p=m\sinh\theta,}
with
\eqn\ren{m=m_0{\gamma e^{\Lambda(1-\gamma)}\over\pi(\gamma-1)}\tan\pi\gamma,
\ \theta=\gamma\lambda,}
with $\gamma\equiv{\pi\over 2\mu}$ and $\lambda$ denotes the bare
rapidity of the particles ($s,a$) (again there are some slight
differences between authors for this formula). As shown in
\Kor\ the $S$-matrix of solitons can be extracted from the Bethe
ansatz equations. The $S$-matrix elements are defined as follows. Suppose
first we have only one soliton, with bare rapidity $\lambda_1$. Then
the total phase shift collected by the wave function when the argument
of the soliton goes around the circle is
\eqn\fiI{\phi_1=m_0L\sinh\lambda_1+\sum_j\phi_0(\lambda_1-\bar{\xi}_j)}
where the sum is taken over all pseudoparticles in the sea and $\bar{\xi}$
indicates shifted  (with respect to the ground state) rapidities due to
 the backflow. Suppose then we have two solitons with bare rapidities
 $\lambda_1$ and $\lambda_2$. Then the total phase shift of the wave
function when the argument of the first soliton again goes around the
 circle is
\eqn\fiII{\phi_2=m_0L\sinh\lambda_1+\sum_j\phi_0(\lambda_1-\tilde{\xi}_j).}
where $\tilde{\xi}$ are shifted rapidities. One then defines the
$S$-matrix element by $\ln S=i(\phi_2-\phi_1)$. Complete computation
shows that it depends only on the difference of the rapidities.
Moreover one also checks that for more particles, the phase shifts
simply add and the scattering can be decomposed as a succession of
two-particle ones.  The resulting $S$-matrix, therefore a solution of
the Yang-Baxter equation, has the well-known matrix elements
\eqn\ssg{a=Z(\theta)\sinh\left({i\pi-\theta\over t}\right),\ b=Z(\theta)
\sinh\left({\theta\over t}\right),\ c=Z(\theta)\sinh\left({i\pi\over t}\right)}
where $a$ corresponds to $ss\rightarrow ss$ scattering, $b$ to
$sa\rightarrow sa$ and $c$ to $sa\rightarrow as$. We give the the
expression for normalization factor $Z$ in sect.\ 7. The symmetry
under $s\leftrightarrow a$ gives the remainder of the elements.  We
have parametrized
\eqn\para{\gamma\equiv{\pi\over 2\mu}\equiv{\pi\over t+1}.}
The $S$-matrix \ssg\ can be manipulated to exhibit $\hat{U}_qsl(2)$
symmetry \ref\BL{D. Bernard and A. Leclair, Nucl. Phys. B340 (1990) 721;
G. Felder and A. LeClair, Int. J. Mod. Phys A7 (1992) 239.} with
\eqn\qq{q=-\exp\left(-{i\pi\over t}\right).}
 This is a dynamical symmetry; there is also a kinematical symmetry
$U_{q_0}sl(2)$ with $q_0=-\exp\left(-{i\pi\over t+1}\right)$
following from the Bethe ansatz equations \ref\PS{V. Pasquier and H.
Saleur, Nucl. Phys. B330 (1990) 523.}. Notice the shift of the
denominator.

In order to reach the deep UV limit, we let the mass $m_0$ go to zero.
Particles with non-vanishing energy must have rapidities with very
large modulus, of the order ${\xi_0}=\ln(M/ m_0)>>1$, where $M$ is
a not-yet-defined parameter with the dimension of mass. There are thus
two regions of interest in which we set respectively
$\xi=\xi_0+\theta$ and $\xi=-\xi_0+\theta$, $\theta$ remaining finite.
The spectrum obviously splits into right and left excitations with
\eqn\elft{E_R\approx{m_0\over 2}e^{\xi_0}e^\theta\equiv{M\over 2}
e^{\theta}=p_R;\qquad E_L\approx{m_0\over2}
e^{\xi_0}e^{-\theta}\equiv{M\over 2} e^{-\theta}=-p_L.}
and we have a {\it doubling} of species $a_{L,R}, s_{L,R}$ (see 1).
For pseudoparticles of the same kind, the phase shifts are unchanged
as $\xi_i-\xi_j=\theta_i-\theta_j$. For particles of different kinds
however $\xi_i-\xi_j \approx\pm 2\xi_0\rightarrow\pm\infty$ so the
phase shifts become constants, independent of the rapidities. As a
result, in the computation of the $S$-matrix, the $LL$ and $RR$
scattering are the {\it same} as the ones for corresponding massive
particles computed above $S_{LL}=S_{RR}=S$\foot{In some cases like in
\ZZI\ there is an additional phase in the definition of $S_{LL}$ and
$S_{RR}$.}, while the $LR$ scattering becomes trivial.

The $S$-matrices in the massless limit can also be obtained by
studying the XXZ spin chain, which is a lattice regularization of the
massless Thirring model. This is a simple generalization of the work
of \FT\ on the XXX chain.

It should be clear finally that the properties of a massless theory
with trivial $LR$ scattering are independent of the mass scale $M$.
Indeed, changing $M$ is equivalent to shifting $\xi_0$, and the analysis only
depends on rapidity differences.

\newsec{Thermodynamic Bethe ansatz}
\nref\YY{C.N. Yang and C.P. Yang, J.Math.  Phys. 10 (1969) 1115.}

The technique we now discuss involves computing
 the free energy of an integrable lattice model (e.g.\
the XXZ model) or an integrable quantum field theory (e.g.\ the
Thirring model) on an infinite line at finite temperature $T$. There
are two approaches to this calculation.  In the traditional ``bare''
approach, one finds the energy and entropy of the states of the model
using the Bethe ansatz. The thermodynamic state is the state which
minimizes the free energy. The limit $T\rightarrow 0$ gives the
ground-state energy and the vicinity $T\approx 0$ the structure of
low-lying excitations. In the massive Thirring model this approach
starts with the bare equations \ba.  In the second approach one
forgets the bare theory completely and studies instead the
thermodynamics of a gas composed of the various ``physical''
excitations (like the soliton and antisoliton of the sine-Gordon
model) scattering with their respective $S$-matrices. As in the first
approach, one determines their energy and entropy and again minimizes
the free energy. The first approach was used in works like
\refs{\YY,\W}. The second approach, pioneered in
\ref\Z{Al.B.Zamolodchikov, Nucl. Phys. B342 (1991) 695.}, is more
recent, and is usually called the
thermodynamic Bethe ansatz (TBA).

The second approach allows some convenient short cuts: instead of
solving the theory one first establishes that it is integrable,
conjectures the excitations and their $S$-matrix using intuition and
symmetry arguments (with some care), and uses the TBA to derive
various properties.  When both approaches can be implemented, they of
course give the same results; the quasiparticle excitations used in
the second can be found from the Bethe ansatz equations by filling the
Fermi or Dirac sea.  However, there are examples (at least in the
massive case) where a lattice model is not integrable but its
continuum limit is an integrable quantum field theory. Usually the
only way of defining the continuum model is by a perturbed conformal
field theory, so the usual Bethe ansatz methods cannot be applied; the
only recourse is to use the second approach. A classic example is the
Ising model at $T=T_c$ in a magnetic field \ZAdv\foot{Observe however
that there is another integrable lattice model based on $E_8$ that is
integrable and has the same scaling limit as Ising in a magnetic
field.}.

As a simple example we describe the TBA for a single type of massless
particle, say right-moving, with energy and momentum parametrized as
in \spec.  The scattering is described by a single $S$-matrix element
$S_{RR}$.  Quantizing a gas of such particles a circle of length $L$
requires the momentum of the $i$th particle to obey
\eqn\quan{\exp\left(i{Me^{\theta_i}\over 2}L\right)\prod_{j\neq i}
S_{RR}(\theta_i-\theta_j)=1. }
One can think of this intuitively as bringing the particle around the
world through the other particles; one obtains a product of
two-particle $S$-matrix elements because the scattering is
factorizable. This is the renormalized equivalent of the bare relation
\ba.

Going to the $L\rightarrow\infty$ limit, we introduce the density of
rapidities indeed occupied by particles $\rho(\theta)$ and the density
of holes $\tilde\rho$. A hole is a state which is allowed by the
quantization condition \quan\ but which is not occupied, so that the
density of possible rapidities is $\rho(\theta)
+\tilde{\rho}(\theta)$.  Taking the derivative of the log of \quan\
yields
\eqn\logquan{2\pi[\rho(\theta)+\tilde{\rho}(\theta)]={ML\over 2}
e^\theta+\int_{-\infty}^\infty \Phi(\theta-\theta')\rho(\theta'),}
where
$$\Phi(\theta)={1\over i}{d\over d\theta}\ln S(\theta).$$
To determine which fraction of the levels is occupied we do the
thermodynamics \YY. The energy is
$${\cal E}=\int_{-\infty}^\infty \rho(\theta){M\over 2}
e^{\theta} d\theta,$$
and the entropy is
$${\cal S}=\int_{-\infty}^\infty\left[
(\rho+\tilde{\rho})\ln (\rho+\tilde{\rho})
-\rho\ln(\rho)-\tilde{\rho}\ln(\tilde{\rho})\right]d\theta.$$
The free energy per unit length ${\cal F}=({\cal E}-T{\cal S})/L$ is found by
minimizing it with respect to $\rho$.  The variations of ${\cal E}$ and
${\cal S}$ are
$$\eqalign{\delta{\cal E}&=\int_{-\infty}^\infty
\delta\rho {M\over 2}e^\t d\t\cr
\delta{\cal S}&=\int_{-\infty}^\infty\left[(\delta\rho+
\delta\tilde{\rho})\ln (\rho+
\tilde{\rho})-\delta\rho\ln(\rho)-\delta\tilde{\rho}\ln(\tilde{\rho})\right]
d\theta.\cr}$$
It is convenient to parametrize
\eqn\param{{\rho(\theta)\over \tilde{\rho}(\theta)}
\equiv\exp\left(-{\epsilon\over T} \right),}
giving
$$\delta {\cal S}=\int_{-\infty}^\infty\left[\delta\rho
\ln\left(1+e^{\epsilon/T}\right)+\delta\tilde{\rho}\ln
\left(1+e^{-\epsilon/T}\right)\right]d\theta.$$
Using \logquan\ allows us to find $\tilde\rho$ in terms of $\rho$.
Denoting convolution by $\star$, this gives $2\pi(\delta\rho+
\delta\tilde{\rho})=\Phi\star\delta\rho$
so
$$\delta {\cal S}=\int_{-\infty}^\infty\left[
{\epsilon\over T}+{\Phi\over 2\pi}\star\ln\left(1+e^{-\epsilon/T}\right)
\right]\delta\rho d\theta.$$
Hence the extremum of ${\cal F}$ occurs for
\eqn\tba{{M\over 2}e^\theta=\epsilon+T{\Phi\over 2\pi}\star\ln\left
(1+e^{-\epsilon/T}\right).}
and one has then, expressing $\tilde{\rho}$ from \logquan\
and using \tba\
\eqn\free{{\cal F}={\cal E}_0-T^2{M\over 4\pi T}\int_{-\infty}^\infty
 e^\theta\ln\left(1+e^{-\epsilon/T}\right)d\theta.}
The ground state energy ${\cal E}_0$ cannot be obtained by this method
since all the information we use is the structure of excitations above
the ground state, so we set ${\cal E}_0=0$ for the rest of this section.

The limit $T\rightarrow 0$ of this system is interesting. We introduce
the positive and negative parts of the pseudoenergy satisfying therefore
\eqn\tzero{{M\over 2}e^\theta=\epsilon^+ +\epsilon^-
-{\Phi\over 2\pi}\star\epsilon^-,}
In this limit the solution is $\epsilon^-=0,\ \epsilon^+={M\over
2}e^\theta$. It follows from this and \param\ that $\rho\rightarrow 0$
as $T\rightarrow 0$, which is required because our TBA provides the
structure of excitations over the ground state.  In general
$-\epsilon^-$ (resp. $\epsilon^+$) gives the excitation energy for
holes (resp. particles).

\nref\BCNA{H.W. Blote, J.L. Cardy and M.P. Nightingale, Phys. Rev. Lett. 56
(1986) 742; I. Affleck, Phys. Rev. Lett. 56 (1980) 746.} The knowledge
of ${\cal F}$ leads to the determination of the central charge of the
theory.  We have ${\cal F}=-{T\over L}\ln Z$, where $Z$ is the
partition function of the one dimensional quantum field theory at
temperature $T$ (in the following we refer to this point of view as
``thermal''). In Euclidean formalism, this corresponds to a theory on
a torus with finite size in time direction $R=1/T$. By modular
invariance, identical results should be obtained if one quantizes the
theory with $R$ as the space coordinate . For large $L$,
$Z=e^{-E(R)L}$, where $E(R)$ is the ground-state (Casimir) energy with
space a circle of length $R$. Thus ${\cal F}=E(R)/R$. In the following
we refer to this as the ``finite-size'' point of view.  Conformal
invariance requires that at a fixed point this Casimir energy is
$E(R)=-{\pi c\over 6R}$, where $c$ is the central charge \BCNA. Going
back to the thermal point of view, ${\cal F}=-{\pi cT^2\over 6}$ and
the specific heat is ${\cal C}=-{\pi cT\over 3}$.


Observe from \tba\ and \free\ that the free energy does not depend on
the mass scale $M$, because it can be rescaled by a shift in
rapidities. By dimensional analysis one has therefore ${\cal
F}=T^2\times\hbox{constant}$.  This scale invariance is a
manifestation of the fact that $S_{LL}$ and $S_{RR}$ are describing
only conformal properties. With massive particles or with nontrivial
left-right massless scattering, ${\cal F}$ does depend on
$M/T$, giving a running central charge.

We can analytically find this central charge from \tba. We take the
derivative of \tba\ with respect to $\theta$ and solve for $e^{\theta}$.
Substituting this in \free, we have
$$\eqalign{{\cal F}&=-{T\over 2\pi}\int d\t
\left[{d\e\over d\t} \ln(1+e^{-\e/T}) -\int d\t' \ln(1+e^{-\e(\t)/T})
\Phi(\t-\t') {d\e\over d\t'}{1\over 1+e^{\e(\t')/T}}\right]\cr
&=-{T\over 2\pi}\int d\t {d\e\over d\t}\left[ \ln(1+e^{-\e/T}) +
(\e-{M\over 2}e^\t){1\over 1+e^{\e(\t)/T}}\right]\cr
&=-{\cal F}-
{T\over 2\pi}\int d\t {d\e\over d\t} \left[\ln(1+e^{-\e/T})+{\e\over
1+e^{\e/T}}\right],\cr }$$
where we use \tba\ again to get to the second line. We can replace the
integral over $\t$ with one over $\epsilon$, giving an ordinary integral
$$
{\cal F}=-{T\over 4\pi}\int_{\e(-\infty)}^{\infty}
 d\e \left[\ln(1+e^{-\e/T})+{\e\over
1+e^{\e(\t)/T}}\right] ,$$
A change of variables gives
\eqn\sIII{{\cal F}=-{T\over 2\pi}L\left({1\over 1+x_0}\right),}
where $L(x)$ is the Rogers dilogarithm function
$$L(x)=-{1\over 2}\int_0^x\left({\ln(1-y)\over y}+
{\ln y\over 1-y}\right) dy,$$
and $x_0\equiv\exp(\ep(-\infty)/T)$ is obtained from \tba\ as
\eqn\xoo{{1\over x_0}=\left(1+{1\over x_0}\right)^I,}
with $I={1\over 2\pi}\int\Phi$.

For example, when the $S$ matrix is a constant, $\Phi=0$, $x_0=1$ and
\eqn\sIIII{{\cal F}=-{T\pi\over 24},}
where we used $L(1/2)={\pi^2\over 12}$.  Here we find $c_L={1\over 4}$.
In a left-right-symmetric quantum field theory, the right sector makes
the same contribution, giving the total central charge $c={1\over 2}$
required for free fermions.

For the nonunitary Lee-Yang $S$-matrix \ref\CM{J. Cardy and G.
Mussardo, Phys. Lett. 225B (1989) 275.} one has $I=-1$.  In that case
$x_0={1+\sqrt{5}\over 2}$. Using $L\left({3-\sqrt{5}\over
2}\right)={2\over 5}{\pi^2\over 6}$ one finds $c_R={2\over 10}$ and
$c={2\over 5}$ after left and right contributions have been collected.
This is indeed the effective central charge for the Lee-Yang problem;
it is not equal to the true central charge $c=-{22\over 5}$ because of
the presence of an operator with negative dimension in the vacuum.

\nref\FI{P. Fendley and K. Intriligator, Nucl. Phys. B372 (1992) 553.}
It is possible to do the same computation with the massless pair
($a,s$) scattering with \ssg. The computation is technically more
complicated because the scattering is non-diagonal. We just refer the
reader to references \refs{\FI,\FS} for details. Simply observe that
in the case $\mu={\pi\over 2}$ the scattering becomes diagonal and
because of the doubling of the number of species, the above
calculation gives rise to $c=1$ as expected.

Besides the central charge, some conformal dimensions can also be
identified using the TBA. To do so one includes an imaginary chemical
potential $\mu_b$ (not to be confused with $\mu$ in section 3) for
each species of particle $b$. As before, we minimize the corresponding
free energy ${\cal G}={\cal E}-T{\cal S}-\sum_b\mu_b{\cal N}_b$, and
find similar results with, in most cases, $\exp(-\epsilon_b/T)$
replaced by $\exp(-(\epsilon_b-\mu_b)/T)$.  In the finite-size point
of view, the introduction of a chemical potential amounts to
considering the theory on a circle of length $R$ with twisted boundary
conditions. As is well known, the ground state energy in that case
gives an effective central charge $c_{eff}=c-24h$ where $h$ is related
to the twist. For $RR$ scattering given by \ssg\ for instance, with
$\mu_{s}=-\mu_{a}=i\alpha\pi T/t$ one finds $h={\alpha^2\over
4t(t+1)}$. Thus one recovers the dimensions of vertex operators in a
Gaussian model.

The question of reconstructing the whole quantum field theory from a
massless scattering theory with no left-right scattering is still
open.  For Virasoro minimal models two independent such
theories are probably necessary, as there are two fundamental quantum groups,
or two labels in the Kac table. A bit of progress in this direction is
presented in the next section.

\newsec{Integrable CFT and massless scattering: Virasoro conserved quantities}
\nref\KN{P.P.Kulish and E.R.
Nisimov, Th. Math. Phys. 29 (1976) 161.}\nref\KF{V. Korepin and L.D.
Faddeev, Th. Math. Phys. 25 (1975) 147.}

We showed in sect.\ 4 how one obtains the free energy in a massless
integrable theory, and compared this result with conformal field
theory predictions.  In an integrable theory, the energy ${\cal
E}\equiv\<E\>$ is just the first of an infinite series of conserved
charges.  These conserved charges can often be expressed as suitably
regularized powers of the energy-momentum tensor. Since we derive the
particle densities of the thermodynamic state, we can calculate the
expectation value of any quantity which can be expressed in terms of
the particles. This suggests that the conserved charges should be
related to expectation values $\<E^n\>$.  In this section we show that
in the sine-Gordon model, this is indeed true, thus verifying
non-perturbatively what had been shown classically and perturbatively
in the quantum theory \refs{\KN,\KF}.

We start with free fermions with antiperiodic boundary conditions on a
circle of length $R$ in order to select the ground state. We are
thinking about the system in the finite-size point of view discussed
in the last section. Consider the quantity
\eqn\ef{\eqalign{\<E^{n}\>_{gs}&={1\over 2}\left({2\pi\over R}\right)^n
\left<\sum_{j=-\infty}^{\infty}(j+1/2)^n:\psi_{-j-1/2}\psi_{j+1/2}
:\right>_{gs}\cr &=(-)^n{1\over 2}\left({2\pi\over
R}\right)^n\sum_{j=0}^ {\infty}(j+1/2)^n.\cr}}
The sum can be evaluated by $\zeta$-function regularization leading to
\eqn\efI{\<E^{2k+1}\>_{gs}={1\over 2}\left({2\pi\over R}\right)^{2k+1}\left
(1-{1\over 2^{2k+1}}\right)\zeta(-2k-1),}
and
\eqn\efII{\<E^{2k}\>_{gs}=0,}
where the last result follows from $\zeta(-2k)=0$. For $k=0$ one gets
${\cal E}=\<E\>=-{\pi\over 6R}{c\over 2}$ with $c=1/2$ ($c/2$ appears because
we concentrate on one chirality).

We can compute the analogous quantity from the thermal point of view
by putting the particles on a circle of large length $L$ at temperature
$T=1/R$. The TBA analysis of sect.\ 4 gives
\eqn\epf{\eqalign{\<E^{n}\>_{TBA}&=\int_{-\infty}^\infty{d\theta}
 \left({Me^\theta\over 2}\right)^{n}\rho(\t)\cr
&=\int_{-\infty}^\infty{d\theta}
  \left({Me^\theta\over 2}\right)^{n}(\rho(\t)+\tilde\rho(\t))
{1\over 1+\exp(\e/T)}\cr
&=nLT\int_{-\infty}^\infty{d\theta\over 2\pi}
  \left({Me^\theta\over 2}\right)^{n}\ln(1+e^{-\e/T}),\cr
}}
where we used the fact that $2\pi(\rho+\tilde\rho)= LTd\e/d\t$, which is
proven by showing that they obey the same integral equation.

For free fermions, $\e=Me^\t/2$.  The expectation value \epf\ is
$$\<E^n\>_{TBA}={L\over R}{n\over 2\pi}R^{-n}\int_0^\infty
x^{n-1}\ln(1+e^{-x})dx,$$
or, restricting to $n=2k+1$
\eqn\epfII{\<E^{2k+1}\>_{TBA}={L\over R}{(2k+1)!\over 2\pi}R^{-2k-1}\left
(1-{1\over 2^{2k+1}}\right)\zeta(2k+2).}
To compare \efI\ and \epfII\ recall the identities
\ref\GR{I. Gradshtein and I. Rishnik, {\it Table of Integrals, Series and
Products} (Academic Press, 1980).}
$$\zeta(2k+2)={(2\pi)^{2k+2}\over 2 (2k+2)!}(-1)^k B_{2k+2};\qquad
\zeta(-2k-1)=-{B_{2k+2}\over 2k+2},$$
where $B_{n}$ are Bernouilli numbers. Hence we find
\eqn\comp{\<E^{2k+1}\>_{TBA}=(-1)^{k+1}{L\over R}\<E^{2k+1}\>_{gs}.}
For $k=0$ we recover ${\cal F}$ as derived in the previous section,
using the fact that ${\cal E}=-{\cal F}$ for theories with no
left-right scattering.  More generally \comp\ follows from the
relation of $\<E^{2k+1}\>_{TBA}$ (resp. $\<E^{2k+1}\>_{gs}$) to the
energy-momentum tensor component $T_{xx}^{k+1}$ (resp. $T_{yy}^{k+1}$)
and $T_{xx}+T_{yy}=0$ at the conformal point.

Notice that when $n$ is even, the results are quite different, since
$\<E^{2k}\>_{TBA}\neq 0$ while \efII\ holds.  The usual argument is
that such even powers do not correspond to any local quantity in the
quantum field theory, and therefore the two results cannot be compared
as we did in \comp.

The generalization of this computation to the case of nontrivial
scattering is not straightforward, but it is useful.  We can compute
-- at least numerically -- the quantities $\<E^{2k+1}\>_{TBA}$ from
\epf.  In general the conformal field theory cannot easily be
described by oscillators as in the free theory above, so we do not
compute the equivalent of $\<E^{2k+1}\>_{gs}$. What we can however
compute using {\it only} the Virasoro algebra are quantities like
\eqn\xx{\<\int :T^{k}:\>_{h},}
where the integral is over the period of the cylinder, and average is
taken in a state of conformal weights $(h,h)$, generalizing the ground
state. The double dots indicate normal ordering on the cylinder.
Recall that this normal ordering leaves room for non-vanishing
expectation values. These can be computed by explicitly performing the
subtraction of the divergent terms. The result coincides with the
simpler zeta regularization.  Such a quantity cannot be directly
compared to $\<E^{2k+1}\>_{TBA}$ because of a non-trivial
renormalization factor. In the free fermion case for instance $\<\int
:T^{2}:\>_{gs}=\left({2\pi\over R}\right)^3{49\over 96}\zeta(-3)$
while $\<E^3\>_{gs}=\left({2\pi\over R}\right)^3{7\over 16}\zeta(-3)$,
because $:T^2:={3\over 8}:\partial^2\psi\partial\psi: -\ {5\over
24}:\partial^3\psi\partial\psi:$.  However this normalization factor
occurs from short-distance singularities and therefore does not depend
on boundary conditions. We can therefore compare ratios of the moments
of $T, \partial T,\dots$ and $\<E^n\>$'s for different boundary
conditions around the circle, or equivalently the choice of state in
which the average \xx\ is taken. As explained in the previous
section, in the thermodynamics this means taking imaginary chemical
potentials for the particles.

A crucial point is that we must treat the theory as a non-minimal,
non-unitary theory with central charge $c=1-{6\over t(t+1)}$. We use
this value of $c$, denoted as $c_{min}$, in the Virasoro algebra
computations. Boundary conditions other than these twisted ones give
rise to $c_{eff}=c_{min}-24h$, where the conformal dimension $h$ is
computed with respect to the $c_{min}$ ground state.  The usual
sine-Gordon ground state (with $c_{eff}=1$) is then interpreted as
arising from an operator of negative dimension. With these definitions
one finds for instance, using the Virasoro algebra and
$\zeta$-function regularization,
\eqn\tI{\<\int :T^{2}:\>_h={1\over 5760}(10c_{eff}^2+40c_{eff}+4c_{min}),}
\eqn\tII{\<\int :T^{3}:\>_h=-{1\over( 24)^3}\left(c_{eff}^3+12c_{eff}^2+
{192\over 5}c_{eff}+{32\over 7}c_{min}+{6\over 5}c_{min}c_{eff}\right),}
\eqn\tIII{\<\int :T\partial^2T:\>_h=-{1\over( 24)^3}\left({48\over 5}c_{eff}+
{32\over 7}c_{min}\right).}
The $c_{eff}$ result from the non-zero $\<L_0\>_h$ on the cylinder.

Our strategy has been simply to compute numerically the values of
$\<E^{2k+1}\>_{TBA}$ for various chemical potentials, denoting
these by $\<E^{2k+1}\>_{\alpha}$.  For sine-Gordon the
chemical potentials are $\mu_{s}=-\mu_{a}=i\alpha\pi T/t$; this
results in $c_{eff}=1-24h$, where $h={\alpha^2\over 4t(t+1)}$.  The
general TBA equations with fugacities are written out in \FS.  Since
the numerics are crucial to obtaining our result, we describe the
methods briefly.  The multi-function generalization of \tba\ is of the
form
$$\epsilon_a (\t) =  \nu_a(\t) + \sum_{b}\int d\t'  \phi_{ab}(\t-\t')
 \ln(1+\lambda_b e^{-\ep_b(\t')/T}) $$
To find $\epsilon_a$ numerically, we solve this iteratively. We guess
the initial $\epsilon_a$; $\ep_a=\nu_a$ usually works. Then one
evaluates the right-hand side numerically by discretizing the integal;
this gives the next guess for $\ep_a$. Usually the iteration
converges; occasionally one needs to use take a linear combination of
$x$(guess)$+(1-x)$(iteration) for the next guess. More elaborate
methods to improve convergence are described in \ref\KM{T.  Klassen
and E. Melzer, Nucl. Phys. B350 (1990) 635.}. Once this procedure has
obtained $\ep_a$ to the desired accuracy, the expression \epf\ for
$\<E^n\>$ can then be numerically evaluated. We note that this
numerical procedure for solving non-linear integral equations is
generally far simpler to implement that those for solving non-linear
differential equations.

The numerical results are rather interesting.  As before, we have
$c_{min}=1-6/t(t+1)$ and $c_{eff}=1-6\alpha^2/t(t+1)$. We find
\eqn\fitt{\eqalign{\<E^3\>_{\alpha}=&f_t(10c_{eff}^2+40c_{eff}+4c_{min})\cr
\<E^5\>_{\alpha}=&g_t(c_{eff}^3+12c_{eff}^2+40c_{eff}+2c_{eff}c_{min}+
{16\over 3}c_{min}+{8\over 21}c_{min}^2).\cr}}
Moreover, we find that at least for $\<E^3\>$, the prefactor takes
reasonably simple values; to excellent numerical accuracy we have
$$f_2={7\over 48}\quad f_3={\pi^2\over 70} \quad f_4={1001\over 7200}\quad
f_5={2\pi^2\over 143}$$
This is a hint that these numbers can be derived analytically from the
TBA, but we have tried and failed to do so. We can make the amusing
observation that $\<E^n\>$ for free fermions can be written in terms
of polylogarithms, just like the dilogarithms written in the last
section for $\<E\>$. We also note that the free energy of the impurity
in the Kondo problem (see \AFL) can be written
as a sum of these conserved quantities, another hint of interesting
hidden structure.

Hence we can fit our numerical results to \tI\ for $n=3$ and to a
linear combination of \tII\ and \tIII\ for $n=5.$
We see that \tI\ is indeed proportional to $\<E^3\>_{TBA}$ and that
\eqn\tIIII{\<E^5\>_\alpha\propto \<\int :T^3:+\left({c_{min}+2\over
12}\right)T\partial^2T:\>_h.}
After the numerical computation was completed we checked that \tIIII\
agrees with the conserved quantity at grade 5 in \ref\SY{R. Sasaki and
I. Yamanaka, Adv. Stud. in Pure Math. 16 (1988) 271.}. This is of
course no surprise. Recall that in the classical sine-Gordon theory,
the conserved quantities are precisely expressed as the sum of odd
powers of the momenta: we simply check here that this result holds in
the quantum theory as well.  This is expected, but as far as we know,
was checked only perturbatively so far \refs{\KN,\KF}. Hence by
massless scattering we recover not only the central charge and
conformal weights of a conformal field theory, but also the conserved
quantities which involve the Virasoro algebra itself, making the
connection between the two points of view a little closer. One might
wonder if there is an action of the Virasoro algebra on the massless
particles.

\newsec{Nontrivial left-right scattering}

The massless Thirring model has no non-trivial $LR$ scattering because
$L$ and $R$ excitations are infinitely separated in the rapidity
plane.  On the other hand, the most interesting situations occur when
the $LR$ scattering is non-trivial. In that case, the theory is not
scale invariant, and is described by two different conformal field
theories in the UV and IR limits. To get such a situation in the
Thirring model, we need to arrange for massless $L$ and $R$
excitations that both occur around the same region of rapidities. A
way to do so is to choose a {\it purely imaginary} bare mass in \thir\
$m_0=-i|m_0|$.  Indeed the result \en\ still holds, so
\eqn\enim{E=-i|m_0|\sum_i\cosh\xi_i.}
If we restrict to the consideration of real energies we need
$Im(\xi)=\pm \pi/2$. Then for $\xi=\pm i{\pi\over 2}+\nu$,
$e=\pm|m_0|\sinh\nu$. One therefore expects the ground state to
resemble figure 2 with the half lines
\eqn\fillgs{Re(\xi)<0,\ Im(\xi)={\pi\over 2};\
 Re(\xi)>0,\ Im(\xi)=-{\pi\over 2},}
filled up. Actually, because the theory is not free, the determination
of the ground state is slightly more delicate --- the interaction
between the various Bethe ansatz roots must be considered.  The choice
of the cutoff is also important, as well as the sign of $g_0$.  One
finds typically that the picture \fillgs\ is almost correct, up to
some exponentially decaying density on the other side of the
half-lines. This produces therefore the necessary massless excitations
around a common region $\xi=0$.  More details of this approach will be
presented in
\ref\SS{S. Skorik and H. Saleur, work in progress.}.

An imaginary mass in the Thirring model is like
imaginary prefactor in front of the cosine term in sine-Gordon, so we
recover the situation discussed in sect.\ 2. As explained there the
appearance of imaginary numbers is more natural than may appear at
first sight. Just as the massive minimal models perturbed by
$\phi_{13}$ are related to the ordinary sine-Gordon model \EY, the
massless flow between minimal models \refs{\ZII,\LC} is related to the
model with imaginary mass. The non-unitarity of the imaginary-mass model
does not exclude unitarity for a subsector (the perturbed massless
minimal model).  For more details see \FSZ.

It is easy to generalize the TBA of section 4 to models with a a
non-trivial $S_{LR}$.  This time, the running central charge depends
nontrivially on $M/T$. Its UV and IR values can be easily found.  As
discussed in the introduction, we expect that the IR conformal field
theory is characterized by only $S_{LL}$ and $S_{RR}$, so its $LR$
scattering should be trivial. One finds as before
\eqn\cir{c_{IR}=c_R+c_L=2c_R={6\over\pi^2}L\left(1\over 1+y_0\right),}
where $y_0$ is the solution of $1/y_0=(1+1/y_0)^{I_1}$ with
$I_1={1\over 2\pi}\int\Phi_{LL}$. In the UV coupling between left and right
particles has to be considered leading to
\eqn\cuv{c_{UV}={6\over\pi^2}\left[2L\left(1\over 1+x_1\right)-
L\left(1\over 1+x_0\right)\right],}
where $1/x_1=(1+1/x_1)^{I_1+I_2}$ and $I_2={1\over 2\pi}\int
\Phi_{LR}$.

A simple example of a massless integrable field theory is the flow
from tricritical to critical Ising model. As discussed in \ZI\ the
spectrum consists of a right mover and a left mover, the Goldstino
resulting from spontaneously-broken supersymmetry. Because the IR
conformal field theory is a free fermion, $S_{LL}$ and $S_{RR}$ must
be trivial. The left-right scattering is given by
\eqn\srl{S_{RL}(\theta_R-\theta_L)=-\tanh \left({\theta_R-\theta_L\over 2}-
i{\pi\over 4}\right).}
The compatibility of left-right and right-left interchange of
arguments of the wavefunction requires that
\eqn\slr{S_{LR}(\theta_L-\theta_R)S_{RL}(\theta_R-\theta_L)=1,}
so here
\eqn\slrI{S_{LR}(\theta_L-\theta_R)=\tanh \left({\theta_L-\theta_R\over 2}-
i{\pi\over 4}\right).}
In the IR limit (the Ising model) where
$\theta_R-\theta_L\rightarrow\infty$ one checks that both matrix
elements go to $-1$ as expected.  With this $S$-matrix we have $I_1=0$
and $I_2=1/2$, so $x_0=1$ and $x_1={\sqrt{5}-1\over 2}$. Using values
of dilogarithms given in sect.\ 4 we find $c_{IR}=1/2$ and
$c_{UV}=7/10$ as desired.

Following \srl\ and \slr\ notice that
\eqn\uni{S_{RL}(\theta)S_{RL}(-\theta)=-1,}
a result that must be carefully compared to the usual
$S(\theta)S(-\theta)=1$ for diagonal massive (or left-left or
right-right) scattering.

\newsec{Sine-Gordon model in a background field}

In this section we discuss the sine-Gordon model in a background field
coupled to the $U(1)$ soliton-number charge. In the traditional bare
approach, this field would modify the Dirac or Fermi sea. In our
approach, this makes it energetically favorable for physical particles
to appear in the vacuum, even at zero temperature. In the sine-Gordon
case with positive background field, only the negatively-charged
particles appear in the vacuum. Their mutual scattering is diagonal,
so the problem is technically easier than the finite-temperature TBA
problem, where both kinds of particles appear in the thermodynamic
state. We discuss both the cases $\lambda$ real and $\lambda$
imaginary, hence giving a (partially) non-perturbative treatment of the
problem raised in sect.\ 2.  For a more complete study see
\FSZ.

\nref\zrefiv{M. Ganin, Izv. Vuzov (Math) 33 (1963) 31.}

The sine-Gordon Hamiltonian with a constant external $U(1)$ gauge
field $A_\mu$ is
\eqn\zi{H=\int dx\left[{1\over 2}\Pi^2+{1\over 2}(\partial\varphi)^2+
\lambda\cos\beta_{SG}\varphi\right] -QA,}
where
\eqn\ziii{
Q=\int j_0 dx={\beta_{SG}\over 2\pi}
\int_{-\infty}^\infty{\partial\varphi\over\partial x}
dx}
In the ordinary case $\lambda$ real, $Q$ is the integer-valued soliton
topological charge, normalized so that the soliton (antisoliton) has
$Q=1$ ($-1$).  $A$ is a constant with the dimension of a length. We
then consider the corresponding specific vacuum energy ${\cal
E}(A,\lambda)$ as a function of $A$.  As before, we parametrize
$\beta_{SG}^2=8\pi{t\over t+1}$.

\nref\BIK{N.M. Bogoliubov, A. Izergin and
V. Korepin, Nucl. Phys. B275 (1986) 687.}
\nref\BIR{N.M. Bogoliubov,
A. Izergin and N. Reshetikhin, J.Phys. A 20 (1987) 5361.}
\nref\WET{F. Woynarovich, H.P. Eckle and T.T. Truong, J.Phys. A22 (1989)
4027.}

Before we turn to the scattering theory, it is worth looking at the action
\zi\ from the perturbative (in $\lambda$) point of view. Dimensional arguments
as well as explicit perturbative calculations show that the background field
works as an infrared  cutoff at scales $\sim A$ and therefore if $A\gg
\lambda^{(1+t)/2}$ the theory is in the ultraviolet regime.  As a leading
$A\to\infty$ approximation we set $\lambda=0$ in \zi.  This theory is
the continuum limit of the XXZ model in a magnetic field, which has
been studied in refs.\ \refs{\BIK-\WET}.  As can easily be inferred
from the action, it is a Gaussian model whose radius of
compactification depends on $A$ (amusing finite-size corrections occur
in the related XXZ model in a field due to commensurability problems
between $R$ and the scale $A$).  Redefining $\del\varphi$ by a shift
gives for the ground-state energy density
\eqn\ziv{{\cal E}(A,0)=-{\beta_{SG}^2\over 8\pi^2}A^2.}
At any critical point, ${\cal E}(A,0)$ is proportional to $A^2$, since
there is no other scale in the problem. The coefficient is
proportional to the chiral anomaly (found from the $J_L J_L$ OPE)
\ref\pkmag{P. Fendley and K. Intriligator, ``Central charges without
finite-size effects'', to appear in Phys. Lett. B.}.

For $\lambda\ne 0$ the scaling argument shows that ${\cal E}(A,\lambda)$ is a
function of the dimensionless variable $$
\xi\equiv\lambda/A^{2/(1+t)}
$$
and by parity has a perturbative expansion in $\xi^2$
\eqn\zvii{
{\cal E}_{\rm pert}(\xi)=-{A^2\over \pi}\sum_{l=0}^\infty k_{2l}\xi^{2l}.}
One can use perturbed conformal field theory to derive
 $$
\eqalign{
k_0=&{t\over 1+t}\cr
k_2=&{\pi^2\over 4}\left({2t\over 1+t}\right)^{2(t-1)/(t+1)}
{\displaystyle{\Gamma\left({1-t\over 1+t}\right)}\over
\displaystyle{\Gamma\left({2t\over 1+t}\right)}.}}
$$
We expect the series \zvii\ to have some finite radius of convergence
$\xi_0$, defining therefore an analytic function ${\cal E}_{\rm pert}
(\xi)$ at $|\xi|<\xi_0$.  The perturbation theory is the same for
$\lambda$ real or imaginary, so \zvii\ holds all around $\xi=0$.
However, the scattering theory depends crucially on the nature of
$\lambda$: for real $\lambda$ the particles are massive, while for
$\lambda$ imaginary they are massless.

Consider first the unitary massive sine-Gordon model ($\lambda$ real
in \zi) and let $m$ be the mass of the corresponding charged particle
(soliton).  As usual the on-mass-shell momenta $(E,p)$ are
parameterized in terms of rapidity $\theta$
$$ E=m\cosh\theta\ ;\ \ \ \ \ \ p=m\sinh\theta $$
In the field \zi\ every soliton (antisoliton)
acquires additional energy $-A$ ($A$).  It is clear that if $A\gg m$ the
state without particles is no longer the ground state. The true vacuum
contains a sea of positively-charged solitons which fill all possible
states inside some ($A$-dependent) ``Fermi interval'' $-B<\theta<B$.
The non-trivial scattering of the solitons certainly
influences the structure of the ground-state sea.  However, only one
kind of particle is in the sea (this can be checked more completely
\refs{\W, \SS}); for $A>0$ this is the soliton. The solitons
scatter diagonally among themselves, the two-particle amplitude being
$a(\theta)$ from \ssg.

As in the finite-temperature case, we define the density of particles
$\rho$ and density of states $\rho +\tilde\rho$.  To obtain the ground
state energy we minimize
\eqn\eqI{{\cal E}^{(Re)}(A)-{\cal E}^{(Re)}(0)= \int (m\cosh\theta-A)
\rho(\theta)d\theta,}
(the superscript $Re$ is added to stress that currently we address the
ordinary sine-Gordon model with real coupling $\lambda$) subject to the
quantization
\eqn\eqII{2\pi[\rho+\tilde{\rho}]=m\cosh\theta+\Phi\star\rho}
where the kernel $\Phi$ follows from the soliton-soliton
$S$-matrix and reads explicitly
\eqn\zxvi{{\Phi(\theta)\over 2\pi}={1\over 2\pi i}{d\over d\theta}\log
 a(\theta)=
\int{e^{i\omega\theta}\sinh{\pi(t-1)\omega\over 2}\over
2\cosh{\pi\omega\over 2}\sinh{\pi t\omega\over 2}}{d\omega\over 2\pi}.}
The equations \eqI\ and \eqII\ and the equations for $B$ (given by
minimizing the energy with respect to $B$) can be put in a more convenient
form by defining the function $f(\t)$ as
\eqn\fore{f (\t )=A - m\cosh \t  +
\int_{-B}^{B} d\t'\Phi(\t -\t')f (\t'),}
where this equation is good only for $|\t|<B$.
Replacing $A-m\cosh\t$ in  \eqI\ with this and using \eqII\ one finds that
\eqn\gsii{{\cal E}(A)= -{m\over 2\pi}\int_{-B}^{B}d\t\ \cosh\t
\ f(\t).}
The boundary conditions $f(\pm B)=0$ determine $B$.

We can understand the meaning of the function $f$ as follows.  Define
$\epsilon^+$ as the energy of particle excitations above the ground
state, and $\epsilon^-$ as the energy of holes. By this definition,
$\e^+\ge 0$ and $\e^-\le 0$.  A variation of the energy is thus
\eqn\eqIII{\delta{\cal E}^{(Re)}(A)
=\int(m\cosh\theta-A)\delta\rho(\theta)d\theta
=\int \epsilon^+\delta\rho-\epsilon^-\delta\tilde{\rho}.}
Using \eqII\ to reexpress $\delta\tilde{\rho}$ as a function of
$\delta\rho$ we find
\eqn\eqIIII{\delta{\cal E}^{(Re)}(A)=\int\left(\epsilon^++\epsilon^--
{\Phi\over 2\pi}
\star\epsilon^-\right)\delta\rho}
so by comparing \eqIII\ and \eqIIII\ we find
\eqn\eqV{m\cosh\theta-A=\epsilon^++\epsilon^--{\Phi\over 2\pi}\star\epsilon^-.}
Using this in \eqI\ gives
\eqn\eqVI{{\cal E}^{(Re)}(A)-{\cal E}^{(Re)}(0)={m\over 2\pi} \int d\t\left[
\cosh\theta\epsilon^- +\epsilon^+\rho-\epsilon^-\tilde{\rho}\right].}
In the ground state $\rho(\t)=0$ when $\epsilon^+(\t)>0$ and
$\tilde{\rho}(\t)=0$ when $\epsilon^-(\t)<0$. This means that the last
two terms vanish, and we obtain \gsii, where $f=-\epsilon^-$.

{}From a formal point of view one may interpret the whole scattering
approach and the resulting system \zxvi--\gsii\ as a way of
summing up the perturbative expansion \zvii\ at real $\xi$ and going
beyond the radius of convergence {\it along} $\lambda$ {\it real}.
One would like a similar tool to sum up \zvii\ at $\lambda$ purely
imaginary. To do so we make a guess inspired by the study of the
massive Thirring model and section 6. We assume that in the
sine-Gordon model with imaginary coupling or Thirring model with
imaginary mass, the number of species doubles, so now we have a
pair of left and right massless particles. We also assume that the $LL$
(identical to $RR$) and the $LR$ scattering are nontrivial. The
energy spectrum \spec\ is gapless. Turn on the background field $A>0$
as before.  The positively charged particles are always excited in the
ground state; we assume as in the massive case that they are the only
particles contributing to the thermodynamics. Now the right- and
left-movers fill respectively the semi-infinite Fermi intervals
$-\infty<\theta<B$ and $-B<\theta<\infty$ with some Fermi boundary
$B\sim \log A/M$. Again it is a straightforward Bethe ansatz exercise
to derive the following system of integral equations
\eqn\zxxix{\eqalign{
QA-{Me^\theta\over 2}&=
f_R(\theta)-\int\limits_{-\infty}^B \Phi_{LL}(\theta-\theta')
f_R(\theta'){d\theta'\over 2\pi}
-\int\limits_{-B}^\infty \Phi_{RL}(\theta-\theta')\rho_L(\theta')
{d\theta'\over 2\pi};\cr
QA-{Me^{-\theta}\over 2}&=
f_L(\theta)-\int\limits_{-B}^\infty \Phi_{LL}(\theta-\theta')
f_L(\theta'){d\theta'\over 2\pi}
-\int\limits_{-\infty}^B \Phi_{RL}(\theta-\theta')f_R(\theta')
{d\theta'\over 2\pi}.}}
The positive functions $f_R(\theta)$ and $f_L(\theta)$ are defined in
the Fermi intervals $\infty<\theta<B$ and $-B<\theta<\infty$ respectively, and
are
restricted by the boundary conditions
\eqn\zxxx{f_R(B)=f_L(-B)=0}
Finally, the ground state energy, which we now call ${\cal E}^{(Im)}(A)$, is
evaluated as follows
\eqn\zxxxii{
{\cal E}^{(Im)}(A)-{\cal E}^{(Im)}(0)=
-{M\over 2\pi}\int_{-\infty}^Be^\theta\ f_R(\theta)d\theta,}
where we have taken into account the obvious symmetry $f_R(\theta)=
f_L(-\theta)$.

In eqn.\ \zxxix\ we introduced the $U(1)$ charges $\pm Q$ of the
massless particles, which, with the knowledge assumed in these
lectures, we cannot fix in advance. It is possible to carry out the
computation with undetermined kernels, and fix them at the end by
requiring the result to provide analytic continuation of the
perturbative series to $\lambda$ imaginary. To save time, let us give
the answer and justify it. One has for $\Phi_{LL}(\theta)$ the {\it same}
expression as in the massive case, but with a {\it shift}
$t\rightarrow t-1$. For $\Phi_{RL}$ we have the same, with a further
shift $\theta\rightarrow
\theta+i{\pi\over 2}(t-1)$. This results in
\eqn\zxxxi{\eqalign{
{\Phi_{LL}(\theta)\over 2\pi}
&={1\over 2\pi i}{d\over d\theta}\log a_{LL}(\theta)=
\int{e^{i\omega\theta}\sinh{\pi(t-2)\omega\over 2}\over
2\cosh{\pi\omega\over 2}\sinh{\pi(t-1)\omega\over 2}}{d\omega\over 2\pi}\cr
{\Phi_{RL}(\theta)\over 2\pi}
&={1\over 2\pi i}{d\over d\theta}\log a_{RL}(\theta)=
-\int{e^{i\omega\theta}\sinh{\pi\omega\over 2}\over
2\cosh{\pi\omega\over 2}\sinh{\pi(t-1)\omega\over 2}}{d\omega\over 2\pi}.\cr}}

Although this new BA system \zxxix--\zxxxii\ has a rather different form from
that of eqns.\ \zxvi--\gsii, it is easy to relate the two in the UV region
$A\to\infty$ where in both systems $B\to\infty$. In this limit the right and
left Fermi intervals have a broad overlap at $-B<\theta<B$. Near say the right
Fermi boundary $\theta\sim B$ (where the main contribution to \zxxxii\ comes
from) we can forget about the left one and solve \zxxix\ for $f_L(\theta)$
by the Fourier transform with $B\to\infty$. The resulting equation for
$f_R(\theta)$ is
\eqn\zxxxiii{rQA-{Me^\theta\over 2}=
f_R(\theta)-\int_{-\infty}^B \Phi(\theta-\theta')f_R(\theta')
{d\theta'\over 2\pi}}
where in terms of the Fourier transforms
\eqn\zxxxiv{
\tilde \Phi(\omega)=\tilde \Phi_{LL}(\omega) +
{[\tilde\Phi_{RL}(\omega)]^2\over 1-\tilde
\Phi_{LL}(\omega)}= {\sinh{\pi(t-1)\omega\over 2}\over
2\cosh{\pi\omega\over 2}\sinh{t\omega\over 2}}}
(compare with eqn.\ \zxvi) and
$$
r=1+{\tilde \Phi_{RL}(0)\over 1-\tilde \Phi_{LL}(0)}={t-1\over t}
$$
It coincides precisely with the corresponding limit $B\to\infty$ of eq. \fore\
provided
\eqn\zxxxvii{
\eqalign{&Q={t\over t-1}\cr
&M=m\cr}}

We pause to discuss the logic. In the UV limit
$\lambda=0$ and we have a free boson after the shift
$\partial\varphi\rightarrow\partial\varphi+A$. We can perturb this
fixed point by $\lambda$ real or $\lambda$ purely imaginary. The first
case is the usual sine-Gordon in the regime where there are only
solitons in the scattering theory. The theory is massive, so the IR
fixed point is the trivial one. The second case is like the
sine-Gordon model with imaginary $\lambda$, where there are only $L$
and $R$ ``solitons''.  As discussed in previous sections the theory is
massless and the IR fixed point is now nontrivial.
Using known results about flow between minimal models (for example the
large-$t$ expansion discussed in sect.\ 2), we expect the IR fixed
point to exhibit the shift $t\rightarrow t-1$.  In the
IR limit, the $LR$ scattering becomes negligible, hence the form of
$LL$ and $RR$ scattering above. For $LR$ scattering we require that in
the opposite UV limit, the two scattering theories look the same,
which is natural since they hold in two regimes connected at the UV
fixed point.

The expansion \zvii\ is obtained by analyzing our two sets of
background field energy equations using a generalized Weiner-Hopf
technique \refs{\zrefiv,\W}. This has been described in detail in
\FSZ, so we do not present the full calculation here.  Instead, we
will explain how to extract the relevant information from the kernels.
The technique relies on the usual Weiner-Hopf trick of dividing the
Fourier transforms of the kernels into a product of two pieces, the
first of which has no poles or zeroes in the lower half plane and the
second none in the upper half plane.  Defining
$1/K_+(\omega)K_-(\omega)\equiv 1-\tilde\Phi(\omega)$, one finds that
expressions of the form
\eqn\contour{\oint
{h(\omega)\over (\omega -i)^2} g(\omega) e^{2i\omega B} {d\omega\over 2\pi
i};\qquad\qquad g(\omega)\equiv{K_+(\omega)\over K_-(\omega)}}
occur regularly in the analysis; the contour covers the upper half
plane. The function $h(\omega)$ is different depending on where we are
in the analysis, but it is analytic in the upper half plane.  First we
discuss the massive case. The poles in the contour are at $\omega=i$
and at the zeros of $1-\tilde\Phi$, which are at $\omega=2in/(t+1)$.
The pole at $\omega=i$ results in the bulk contribution ${\cal E}(0)$.
Ignoring the bulk term, \contour\ can be written as a series in
$\exp(-4B/(t+1))$. In particular, the boundary condition results in an
equation
$${M\over A}e^B = const  + \sum_n h_n g_n e^{-4nB/(t+1)}$$
where $h_n$ and $g_n$ are the residues of $h(\omega)/(\omega-i)^2$ and
$g(\omega)$, respectively. The $h_n$ themselves also obey an equation of this
form, so for large $A/M$, we can write $e^B$ and $h_n$ each as a series in
$(A/M)^{-4/(t+1)}$. The energy is also given by a term like \contour, so it
too must be a series in $(A/M)^{-4/(t+1)}$ as in \zvii:
\eqn\ereal{
{\cal E}^{Re}({A,M})={\cal E}^{(Re)}(0) -{A^2\over\pi}\sum_{n=0}^\infty
k_{n}\left({M\over A}\right)^{4n/(t+1)}.}

This gives the result \zvii\ for ordinary sine-Gordon. For the
imaginary coupling, we must first rewrite the equations \zxxix\ in
Weiner-Hopf form.  The general result is that
$$ \eqalign{A\to&Q (1+{\tilde\Phi_{LR}(0)\over 1-\tilde\Phi_{LL}(0)})A
\cr
{1\over K_+(\omega)K_-(\omega)}=&1-\tilde\Phi_{LL}-{\tilde\Phi_{LR}^2\over
1-\tilde\Phi_{LL}}\cr g(\omega)=&{K_+(\omega)\over K_-(\omega)}
{\tilde\Phi_{LR}\over 1-\tilde\Phi_{LL}}.\cr}$$
The $K_+$ and $K_-$ obtained for the massless case are exactly the
same as the ones obtained above for the massive one. The extra piece
in the above expression for $g(\omega)$ is
$-\sinh(\pi\omega/2)/\sinh(\pi t\omega /2)$ here.  It results in no
other additional poles in the contour because of the zeros in $g$.
Its only effect is to change $g_n$ to $(-1)^n g_n$ (again ignoring the
bulk piece).  Since the $h_n$ above are not changed, we then find that
the series for the massless flow is exactly the same as in ordinary
sine-Gordon, except that the signs of every other term are different:
$$ {\cal E}^{(Im)}(A,M)= {\cal E}^{(Im)}(0)-{A^2\over \pi}
\sum_{n=0}^\infty (-)^nk_n\ \left({M\over rQA}\right)^{4n/(1+t)}
$$
with precisely the same coefficients $k_n$ as in expansion \ereal.

We conclude that up to the known bulk vacuum energy contributions the
massless BA system \zxxix--\zxxxii\ gives the correct analytic
continuation of the massive one \zxvi--\gsii\ to purely imaginary
$\xi$, providing \zxxxvii\ holds.  In particular, the low-temperature
mass scale $M$ is equal to the high-temperature scale $m$.

This entire discussion can presumably be put on firmer ground by solving the
Thirring model with imaginary mass using the traditional Bethe ansatz
approach \SS.

\newsec{Conclusions}

The examples presented here have mainly been the sine-Gordon and
Thirring models, with and without a background field. These
calculations can be extended to the truncated (RSOS) cases, the latter
situation being completely unitary.  See the references
\refs{\FSZ,\RS}.

It seems that the physics of massless flows is intimately related with
the one of {\it symmetry breaking} and that the conformally-invariant
IR fixed points are generally some sort of Goldstone phase.  Two of
the simplest cases, the flow from tricritical to critical Ising and
the flow from dilute to dense polymers have to do respectively with
$N=1$ and $N=2$ spontaneous SUSY breaking (the latter being possible
because of the non-unitarity). Moreover, as already mentioned, the
sine-Gordon model with imaginary coupling describes the flow from
critical to low temperature $O(n)$ model (with $n=2\cos{\pi\over t}$).
The Mermin-Wagner theorem preventing spontaneous breaking of
continuous symmetry does not apply to the cases $n$ non-integer. As a
consequence, $O(n)$ models in two dimensions $-2<n<2$ have a
low-temperature phase which is massless and has properties reminiscent
of a Goldstone phase \ref\Sa{H. Saleur, Phys. Rev. B35 (1987) 3657.}
(for instance they qualitatively agree with what can be deduced from
the $\epsilon$-expansion in higher dimensions, extended formally to
$D=2$).

Some important questions remain to be addressed. For instance, can one
seriously describe conformal field theories using massless scattering
(reconstruct Green functions using form factors)?  As explained above,
massless scattering is a sort of perturbation of the IR fixed point.
Can one, using it (i.e.\ probably using the conserved quantities) give
a meaning to the conformal perturbation theory of an IR fixed point by
an irrelevant operator?
\vskip2cm
{\bf Acknowledgments}: These lectures were presented by H.S. who
thanks the organizers and the students at Trieste for many interesting
questions.  We have benefited a great deal by interacting with our
collaborators K.\ Intriligator, N.\ Reshetikhin, S.\ Skorik and Al.B.\
Zamolodchikov.  P.F.\ and H.S.\ were supported by the Packard
foundation and DOE grant No. DE-FG03-84ER40618.
\vfill
\eject
{\bf Figure  Captions}
\bigskip
\noindent Figure 1:
The structure of the ground state in the massive Thirring model. Left
and right massless excitations are observed in the limit
$\xi\rightarrow\pm\infty$. For instance, a pair of left and right
solitons is obtained by making two holes for $|\xi|\gg 1$.
\smallskip
\noindent Figure 2:
Schematic structure of the ground state for imaginary $m_0$.

\listrefs
\bye